\documentstyle[preprint,prb,aps,amsfonts,epsf]{revtex}
\begin{document}
\title{Non-Abelian bosonization of the frustrated antiferromagnetic spin-1/2
chain}
\author{Dave Allen and David S\'en\'echal}
\address{D\'epartement de Physique and Centre de Recherche en Physique du
Solide,}
\address{Universit\'e de Sherbrooke, Sherbrooke, Qu\'ebec, Canada J1K 2R1.}
\address{\tt CRPS-96-11 [cond-mat/9606007]}
\date{May 1996}
\maketitle

\baselineskip=14pt plus 0.5pt
\begin{abstract}
We study the spin-1/2 chain with nearest neighbor ($\kappa_1$) and
next-nearest neighbor ($\kappa_2$) interactions in the regime $\kappa_2\gg
\kappa_1$, which is equivalent to two chains with a `zig-zag' interaction.
In the continuum limit, this system is described in term of two coupled level-1
WZW field theories. We illustrate its equivalence with four off-critical Ising
models (Majorana fermions). This description is used to investigate the opening
of a gap as a function of $\kappa_1$ and the associated spontaneous breakdown
of parity. We calculate the dynamic spin structure factor near the wavevectors
accessible to the continuum limit. We comment on the nonzero string order
parameter and show the presence of a hidden  ${\Bbb Z}_2\times{\Bbb  Z}_2$
symmetry via a nonlocal transformation on the microscopic Hamiltonian. For a
ferromagnetic interchain coupling, the model is conjectured to be critical,
with different velocities for the spin singlet and spin triplet excitations.
\end{abstract}

\section{Introduction}
One-dimensional quantum antiferromagnets have peculiar properties (exotic
ground states, gapped excitations, etc.) which are not accessible to
traditional methods like spin-wave theory or perturbation theory, but require
the use of variational, numerical, or field-theoretical approaches. In
particular, field-theoretical methods have been used successfully to predict
the existence of an excitation gap in the spin-1 Heisenberg
chain\cite{Haldane83} and the scaling behavior of the spin-1/2 Heisenberg
chain.\cite{Affleck86,Affleck89}. In the latter case, Witten's non-Abelian
bosonization\cite{Witten84} was used to express the spin-$\frac12$ Heisenberg
chain as a Wess-Zumino-Witten model perturbed by irrelevant interactions.

In this work, we apply non-Abelian bosonization to the spin-$\frac12$
Heisenberg chain with nearest-neighbor (NN) coupling $\kappa_1$ and
next-nearest-neighbor (NNN) coupling $\kappa_2$ in the regime
$\kappa_2\gg \kappa_1$. This system may also be viewed as two spin-$\frac12$
chains coupled with a `zig-zag' interaction $\kappa_1$ (see figure). This
latter representation makes sense physically, since such arrangements of atoms
occur frequently (see, for instance, Ref.~\onlinecite{Matsuda95}). The
Hamiltonian of this system is
\begin{equation}
\label{hamil1}
H=\kappa_1\sum_i {\bf S}_i\cdot {\bf S}_{i+1}+
\kappa_2\sum_i{\bf S}_i\cdot{\bf S}_{i+2}
\end{equation}
wherein the spins are indexed consecutively along the zig-zag.
This model has recently been studied by White and Affleck.\cite{White96}
We shall extend the somewhat brief theoretical analysis of
Ref.~\onlinecite{White96} and describe the system in terms of four massive
real fermions (or Ising models). This will allow for an easy calculation of
the dynamic spin structure factor $S(q,\omega)$. We will also discuss the
occurrence of a string (topological) order parameter and the associated
${\Bbb Z}_2\times{\Bbb Z}_2$ symmetry. We shall take some time to describe in
sufficient detail the correspondence between the system of two spin-$\frac12$
chains  and four real fermions, since this is unfamiliar ground for many.

Let us summarize here the main results of this paper and explain its
organization. In Sect.~\ref{NAbosS} we set up the description of two coupled
spin-$\frac12$ chains in terms of four Majorana fermions. This implies a quick
review of the non-Abelian bosonization of a single spin chain (its
representation as a level-1 WZW model). The main result of
this section is the representation
$(\ref{J:4fer},\ref{g:4fer})$ of the fields of the level-1 WZW model in terms
of four Majorana fermions and their associated order and disorder fields. This
representation allows for a representation of the spin operator and interchain
interaction with the help of Eq.~(\ref{spinJJ}). In Sect.~\ref{RGS}, we write
the interchain interaction (and the marginal intrachain perturbation) in terms
of the four fermions and discuss the renormalization of the couplings and
velocities. A mass scale $m \sim \kappa_2\exp-(\kappa_2/\kappa_1)$ appears
dynamically and provides a mass for the fermions, accompanied by a spontaneous
breaking of parity. Lorentz invariance is explicitly broken by the interchain
interaction and one of the four fermions acquires a distinct mass and velocity.
In Sect.~\ref{correlS} we set up the calculation of the spin structure factor
$S(q,\omega)$ (the imaginary part of the dynamic spin susceptibility) near the
four wavevectors available to the continuum limit: $q=0,\pi,\pm\pi/2$ (when
considering wavevectors, we regard the system (\ref{hamil1}) as a single,
frustrated chain and not as two coupled chains). The single-spin excitations
appear at a frequency $\omega\sim m$ near $q=\pm\pi/2$ while a two-particle
continuum appears near $q=0$ and $q=\pi$, like for the spin-ladder. 
In Sect.~\ref{stringS} we show that the nonlocal string order parameter
of Ref.~\onlinecite{Nijs89} is nonzero in the ground state, and how this
breaks down a hidden ${\Bbb Z}_2\times{\Bbb Z}_2$ symmetry of the system.
We also perform an exact (i.e., discrete) nonlocal transformation of
(\ref{hamil1}) that reveals this ${\Bbb Z}_2\times{\Bbb Z}_2$ symmetry.
In Sect.~\ref{discussionS} we discuss the difference between this system
and the usual spin ladder and address the case of ferromagnetic interchain
coupling. A quick, largely notational review of WZW models and of the Ising
model (Majorana fermion) is given in Appendices A and B.

\section{Continuum description of two spin chains}
\label{NAbosS}
%
\subsection{Non-Abelian bosonization}

From the Bethe Ansatz solution we know that the spin-$\frac12$ Heisenberg
chain Hamiltonian
\begin{equation}
\label{Heisenberg}
H = \kappa\sum_i {\bf S}_i\cdot {\bf S}_{i+1}
\end{equation}
is critical. It was also argued by Affleck\cite{Affleck86} that this critical
point is well described by a level-1 Wess-Zumino-Witten conformal field
theory\cite{noteA} (cf. Appendix A). This equivalence is demonstrated by
starting from the half-filled Hubbard model with hopping integral $t$ and
on-site repulsion $U$ and taking the continuum limit. The charge degrees of
freedom are then described by a Bose field $\varphi$ which becomes massive
for arbitrary small $U$, while the spin degrees of freedom are described by
the level-1 WZW model. At $U=0$ the characteristic velocity $v$ of the WZW
model is simply the Fermi velocity $v_F=|t|a_0$ ($a_0$ is the lattice spacing).
For $U>0$ the velocity $v$ of the spin degrees of freedom is renormalized
by $U$ and differs from the velocity of the charge excitations (spin-charge
separation). Moreover, the continuum limit of the Hubbard Lagrangian contains
an additional term:
\begin{equation}
\label{marginal1}
{\cal L}_1 = -\lambda J^a \bar J^a
\end{equation}
where $J^a$ and $\bar J^a$ ($a=1,2,3$) are the left and right components of the
$SU(2)$ currents of the level-1 WZW model and $\lambda\sim U/|t|$ (we will work
in the imaginary-time Lagrangian formalism; $\cal L$ denotes the Lagrangian
density). This perturbation is marginally irrelevant. Thus, at long distances,
the spin degrees of freedom are exactly described by the level-1 WZW
model.\cite{Affleck86}

Additional perturbations to the Heisenberg Hamiltonian (\ref{Heisenberg}) may
be expressed in terms of WZW fields by using the following continuum-limit
expression for the spin operator ${\bf S}_i$:\cite{noteB}
\begin{equation}
\label{spinJJ}
{S^a(x)\over a_0}
={1\over 2\pi}\left( J^a(x)+\bar J^a(x)\right)
+(-1)^{x/a_0}\Theta{\rm
Tr}(g(x)\tau^a)
\end{equation}
where $\tau^a$ are the usual Pauli matrices and $g(x)$ the fundamental WZW
field (an $SU(2)$ matrix). The factor $(-1)^{x/a_0}$
alternates from one site to the next and $\Theta$ is a nonuniversal constant.
The first two terms of Eq. (\ref{spinJJ}) constitute the local magnetization
and the last term is the local staggered magnetization.

Let us now turn our attention to the system (\ref{hamil1}). In the regime
$\kappa_1\ll \kappa_2$ and in the continuum limit, it may be regarded as two
level-1 WZW models, plus some perturbations. Let $J^a$ and $\bar J^a$ denote
the $SU(2)$ currents on one chain and $J'^a$ and $\bar J'^a$ the corresponding
currents on the other chain. The first perturbation is
marginally irrelevant and given by two copies of (\ref{marginal1}):
\begin{equation}
\label{inter2}
{\cal L}_2 = -2\lambda_2 (J^a \bar J^a + J'^a \bar J'^a)
\end{equation}
where $\lambda_2\sim U/|t|$. The second perturbation is the interchain
interaction ($\kappa_1$). In the continuum limit and using Eq. (\ref{spinJJ}),
it can be shown without difficulty to be
\begin{equation}
\label{inter1}
{\cal L}_1 = 2\lambda_1 (J^a+\bar J^a) (J'^a+\bar J'^a)
\end{equation}
where $\lambda_1\sim \kappa_1/|t|>0$.

The relevance or irrelevance of a perturbation is determined, as a first
approximation, from the scaling dimensions of the various fields at the WZW
fixed point. In terms the conformal dimensions $(h,\bar h)$ appearing in
(\ref{OPE:T-phi}), the scaling dimension is $\Delta=h+\bar h$ and the
planar spin is $h-\bar h$. Since the conformal dimensions of $J^a$ and
$\bar J^a$ are respectively $(1,0)$ and $(0,1)$, a perturbation of the
form $J^a\bar J^a$ (like (\ref{inter2})) is marginal, while a perturbation of
the form $J^aJ^a$ violates Lorentz (or rotation) invariance. In fact, it
renormalizes the characteristic velocity of the theory (see below). The
interaction (\ref{inter1}) is marginal, except for a velocity renormalization.

WZW models, although they possess conformal invariance, are not easy to
deal with, especially in what regards the calculation of correlation functions.
In some cases (i.e., for some values of the level $k$) the WZW model is
equivalent to a theory of free fields. Then the calculation of correlation
functions becomes an almost trivial task and the overall analysis is much
simplified, in particular the study of the vicinity of the fixed point.
Such a free-field description is possible
in the case of two coupled level-1 WZW models: two such
models are equivalent to one level-2 WZW model, plus one Ising
model (or real fermion, see appendix B). This equivalence was already used in
Ref.~\onlinecite{Totsuka95} to describe the spin ladder with bond alternation.
Moreover, the level-2 WZW model is equivalent to three
Ising models.\cite{Fateev86} We thus have three different ways of describing
the system (\ref{hamil1}) in the continuum limit:
\begin{mathletters}
\label{3ways}
\begin{eqnarray}
&{\rm WZW}_{k=1}\otimes {\rm WZW}_{k=1} \\
&{\rm WZW}_{k=2}\otimes {\rm Ising} \\
&({\rm Ising})^4
\end{eqnarray}
\end{mathletters}%
The representation (\ref{3ways}b) may be useful from the point of view of
symmetry since the interacting terms break down the $SU(2)\times SU(2)$
symmetry to $SU(2)$. However, the representation (\ref{3ways}c) is more
practical for calculations since it is made entirely of free fields and its
off-critical ($\kappa_1\neq0$) behavior may be characterized by ordinary
fermion mass terms. All is not trivial, however, since the Ising model
contains order and disorder fields in addition to a real fermion field, and
these three fields cannot be expressed locally in terms of each other.
Nevertheless, their correlation functions are known. An additional difficulty
comes from the breaking of Lorentz invariance by the perturbation
(\ref{inter1}).

We identify operators in two different languages of (\ref{3ways}) by
requiring their operator product expansions (OPE) to be compatible. The OPE for
the WZW models and the Ising model are given in appendices A and B. The
correspondence of operators belonging to the pictures (\ref{3ways}a) and
(\ref{3ways}c) is the object of the next subsection.

\subsection{Description in terms of four Ising models}
The WZW$_{k=1}$ model cannot be simply represented in terms of two Majorana
fermions, even if the central charge is the same in both cases ($c=1$). The
reason is the nonexistence of a real, spin-$\frac12$ representation of $SU(2)$.
However, two copies of WZW$_{k=1}$ is equivalent to an $SO(4)$ WZW model, and
the latter group admits a representation in terms of four real fermions. A
representation of the WZW currents $J$ and $J'$ in terms of four
Majorana fermions $\psi_{1,2,3,0}$ follows immediately and its structure bears
a strong resemblance with the chiral generators of the Lorentz group:
\begin{mathletters}
\label{J:4fer}
\begin{eqnarray}
J^1 &=  \frac12 i(\psi_1\psi_0-\psi_2\psi_3) \\  
J^2 &=  \frac12 i(\psi_2\psi_0-\psi_3\psi_1) \\  
J^3 &=  \frac12 i(\psi_3\psi_0-\psi_1\psi_2)
\end{eqnarray}
\end{mathletters}%
(the corresponding expressions for $J'^a$ are obtained by reversing the sign
of $\psi_0$). Using the OPE's (\ref{OPE:fermion}) and Wick's theorem, it is
a simple task to check that the OPE's (\ref{OPE:JJ}) are satisfied.

A representation of the matrix fields $g$ and $g'$ (the staggered
magnetizations of the two chains) in terms of Ising fields is also needed in
order to calculate correlation functions, and may be found in the following
fashion. First, since
$g$ and $g'$ have conformal dimensions
$(\frac14,\frac14)$, they must be products of four order and disorder fields,
such as $\sigma_1\sigma_2\sigma_3\sigma_0$,
$\sigma_1\mu_2\sigma_3\mu_0$, and so on (there are $2^4=16$ such products).
Second, the action of each of the currents $J^a$, $\bar J^a$, $J'^a$,
$\bar J'^a$ may be calculated on these 16 products, using the OPE's
(\ref{OPE:Ising}e-h). The result is a 16-dimensional matrix representation of
the currents. According to the OPE (\ref{OPE:J-g}), the field
$g_{\frac12\frac12}$ is an eigenvector of $J^3$ with eigenvalue $-\frac12$.
Once such an eigenvector is found, one may apply on it the other components of
the currents $J$ and $\bar J$ and thus obtain the other components of $g$. Only
one eigenvector allows a nontrivial solution (i.e., nonzero values of all the
components of $g$). The same procedure is used for $g'$, with the currents
$J',\bar J'$. At last, one finds the following representation (we used the
decomposition (\ref{decomp})):
\begin{equation}
\label{g:4fer}
\begin{array}{ll}
g_0 &= \sigma_1\sigma_2\sigma_3\sigma_0 - \mu_1\mu_2\mu_3\mu_0 \\ 
g_1 &= \mu_1\sigma_2\sigma_3\mu_0 - \sigma_1\mu_2\mu_3\sigma_0 \\
g_2 &= \sigma_1\mu_2\sigma_3\mu_0 + \mu_1\sigma_2\mu_3\sigma_0 \\
g_3 &= \sigma_1\sigma_2\mu_3\mu_0 - \mu_1\mu_2\sigma_3\sigma_0 
\end{array}\qquad\qquad
\begin{array}{ll}
g'_0 &= \phantom{-}\sigma_1\sigma_2\sigma_3\sigma_0 + \mu_1\mu_2\mu_3\mu_0 \\
g'_1 &= -\mu_1\sigma_2\sigma_3\mu_0 - \sigma_1\mu_2\mu_3\sigma_0 \\
g'_2 &= -\sigma_1\mu_2\sigma_3\mu_0 + \mu_1\sigma_2\mu_3\sigma_0 \\
g'_3 &= -\sigma_1\sigma_2\mu_3\mu_0 - \mu_1\mu_2\sigma_3\sigma_0
\end{array}
\end{equation}
Note that
the OPE's $J^a(z)g'_i(w,\bar w)\sim0$ and $J'^a(z)g_i(w,\bar w)\sim0$ are
satisfied, as they should: the two chains are independent at this stage.

It is also possible to calculate the OPE of $g$ with itself, with the help of
Eqs~(\ref{OPE:Ising}a-d). This is a bit tricky, since one must remember to
anticommute the different disorder fields. With the normalization chosen above
and omitting terms that do not diverge as $z\to w$, the end result coincides
with Eq.~(\ref{OPE:gg}) for $g$ and $g'$, plus the OPE $g_i(z,\bar
z)g'_j(w,\bar w)\sim 0$. Thus, the representation (\ref{J:4fer},\ref{g:4fer})
is a complete and faithful representation of two independent copies of the
WZW$_{k=1}$ model.

\section{Velocity renormalization and RG analysis}
\label{RGS}
We are now able to write down the Lagrangian associated to the continuum
limit of (\ref{hamil1}) solely in terms of real fermions. The noninteracting
part ${\cal L}_0$, equivalent to the two level-1 WZW models, is the
free-fermion Lagrangian:
\begin{equation}
\label{freeham}
{\cal L}_0 = {1\over 2\pi}
\sum_{i=0}^3 v_i(\psi_i\bar\partial\psi_i +
\bar\psi_i\partial\bar\psi_i)
\end{equation}
where $v_0=\cdots=v_3=v$ is the velocity of spin excitation in isolated chains.
The $2\pi$ factor in (\ref{freeham}) is needed for consistency with the OPE
(\ref{OPE:fermion}). 

The interacting terms (\ref{inter2},\ref{inter1}) may be expressed in terms of
the following operators:
\begin{mathletters}
\label{opers12}
\begin{eqnarray}
O_1 &=& \psi_1\bar\psi_1\psi_2\bar\psi_2 + \psi_1\bar\psi_1\psi_3\bar\psi_3
+ \psi_2\bar\psi_2\psi_3\bar\psi_3 \\
O_2 &=& \psi_0\bar\psi_0(\psi_1\bar\psi_1+\psi_2\bar\psi_2+\psi_3\bar\psi_3)
\end{eqnarray}
\end{mathletters}%
The interaction (\ref{inter2}) is simply
\begin{equation}
{\cal L}_2 = -\lambda_2 (O_1+O_2)
\end{equation}
The translation of (\ref{inter1}) requires more care, however, since it
implies regularized products of identical fermions. The following OPE must
be used to extract the regular terms:
\begin{eqnarray}
\psi_i(z)\psi_j(w) &=& \delta_{ij}\left\{
{1\over z-w} + 2(z-w)T^{(i)}(w) + \cdots\right\}\cr
\bar\psi_i(\bar z)\bar\psi_j(\bar w) &=& \delta_{ij}\left\{
{1\over \bar z-\bar w} 
+ 2(\bar z-\bar w)\bar T^{(i)}(\bar w) + \cdots\right\}
\end{eqnarray}
where $(T^{(i)},\bar T^{(i)})$ is the energy-momentum tensor
(\ref{EMtensor}) of the $i$-th Ising model. The result is
\begin{equation}
\label{inter1B}
{\cal L}_1 = \lambda_1 (O_1-O_2) + \lambda_1\left[
-3(T^{(0)}+\bar T^{(0)}) + \sum_{i=1}^3 (T^{(i)}+\bar T^{(i)})\right]
\end{equation}
Apart from the energy-momentum terms, this interaction coincides with the
marginal interchain interaction obtained by Shelton and al.\cite{Shelton96}
using Abelian bosonization. The effect of the energy-momentum tensor is simply
to renormalize the speeds $v_i$ of the fermions. Indeed, consider the
Lagrangian (we have restored the velocity $v$ in the interaction term)
\begin{equation}
{\cal L} = {v\over2\pi}(\psi\bar\partial\psi +\bar\psi\partial\bar\psi)
-\frac12 v\lambda(\psi\partial\psi+\bar\psi\bar\partial\bar\psi)
\end{equation}
(cf. Eq.~(\ref{EMtensor})) where $\lambda$ is a dimensionless parameter.
One may combine the energy-momentum tensor with the kinetic term and
this amounts to the following renormalizations of the speed and fields:
\begin{equation}
\label{vrenorm}
v \to v{1-\pi\lambda\over 1+\pi\lambda}\qquad\qquad
(\psi,\bar\psi)\to {1\over\sqrt{1+\pi\lambda}} (\psi,\bar\psi)
\end{equation}
In the case at hand, the velocity $v_0$ is renormalized differently from
$v_1,v_2,v_3$:
\begin{equation}
\label{vrenorm2}
v_0 \to v_0{1+3\pi\lambda_1\over 1-3\pi\lambda_1}\qquad
v_i \to v_i{1-\pi\lambda_1\over 1+\pi\lambda_1}\qquad(i=1,2,3)
\end{equation}
After the field renormalization, the interaction Lagrangian takes the
following form:
\begin{equation}
\label{interac}
{\cal L}_{\rm int.} = {\lambda_1-\lambda_2\over (1+\pi\lambda_1)^2}O_1
-{\lambda_1+\lambda_2\over (1+\pi\lambda_1)(1-3\pi\lambda_1)}O_2
\end{equation}
The $O(4)$ symmetry of the fixed-point, obvious in the
Lagrangian (\ref{freeham}), is violated by the interchain coupling
$\lambda_1$, both in the interaction (\ref{interac})
and by the distinct renormalization of $v_0$.

The interaction terms (\ref{interac}) are marginal, since they have conformal
dimensions $(1,1)$. Their behavior under renormalization-group flow is
characterized by their $\beta$-functions. Instead of calculating the latter in
the usual way (a one-loop Feynman diagram), let us follow
Polyakov,\cite{Polyakov72} who has shown that the $\beta$-functions of a
critical system perturbed by marginal terms are related to the coefficients of
the operator algebra. Explicitly, consider the perturbed action
\begin{equation}
S=S_0+\sum_i\lambda_i\int d^2x\; \phi_i(x)
\end{equation}
where $S_0$ is the fixed-point action and the $\phi_i(x)$ are marginal
operators ($h=\bar h=1$). Let the operator algebra be of the form:
\begin{equation}
\phi_i(x)\phi_j(y)\sim{C_{ijk}\phi_k(y)\over\vert x-y\vert^2}
\end{equation}
Then the renormalization-group flow of the couplings $\lambda_i$, characterized
by the $\beta$-functions $\beta_{ijk}(L)$, is (in Euclidian space-time)
\begin{equation}
\label{RG1}
{d\lambda_i\over d{\rm ln}L}=\beta_{ijk}\lambda_j\lambda_k=
-2\pi C_{ijk}\lambda_j\lambda_k
\end{equation}

If we apply this method for the perturbation (\ref{interac}), we
must use the OPE
\begin{equation}
(\psi_i\bar\psi_i)(z)(\psi_j\bar\psi_j)(w) \sim -{\delta_{ij}\over|z-w|^2}
\end{equation}
and realize that the eigenmodes of the RG flow are the operators
\begin{equation}
{\cal K}_+ = O_1+O_2 \qquad {\cal K}_- = O_1-O_2
\end{equation}
which have the OPE
\begin{eqnarray}
\label{OPE:Kpm}
{\cal K}_\pm(z){\cal K}_\pm(w) &\sim&
{6\over|z-w|^4} - {4\over|z-w|^2}{\cal K}_\pm(w)
+ {1\over|z-w|^2}{\cal O}(T+\bar T) + \cdots \cr
{\cal K}_+(z){\cal K}_-(w) &\sim&
{1\over|z-w|^2}{\cal O}(T+\bar T) + \cdots 
\end{eqnarray}
where ${\cal O}(T+\bar T)$ stands for terms containing the energy-momentum
tensor which, although they have the right scaling dimensions, also have
nonzero planar spin and do not contribute to the beta functions.  The terms
in $|z-w|^{-4}$ only contribute to a shift of the vacuum energy and will
be ignored. The interaction (\ref{interac}) may be expressed as a linear
combination of the operators ${\cal K}_\pm$: 
\begin{equation}
\label{interac2}
{\cal L}_{\rm int.} = \lambda_+{\cal K}_+ + \lambda_-{\cal K}_-
\end{equation}
with
\begin{equation}
\lambda_\pm = {1\over2(1+\pi\lambda_1)}\left\{
\mp{\lambda_1+\lambda_2\over 1-3\pi\lambda_1}
+{\lambda_1-\lambda_2\over 1+\pi\lambda_1}\right\}
\end{equation}
The RG equations obtained from (\ref{RG1}) and (\ref{OPE:Kpm}) are
\begin{equation}
\label{RGflow}
{d\lambda_\pm\over d\ln L} = 8\pi \lambda_\pm^2 
\end{equation}
If $\kappa_1\ll\kappa_2$ the starting point is $\lambda_1\ll1$
and $\lambda_2\sim 1$, thus $\lambda_+$ is negative and $\lambda_-$
positive and small. Under this flow, $\lambda_+$ renormalizes to zero
(it is marginally irrelevant, like for an isolated spin chain) and
$\lambda_-$ is marginally relevant. By following the RG flow until $L=\xi$ (the
correlation length), we conclude that $\xi\sim\exp(1/\lambda_-)$: a dynamical
length scale $\xi$ has set in. If we concentrate on the Heisenberg model
(\ref{hamil1}) without referring to the underlying Hubbard model, the
characteristic spin velocity should be $v\sim \kappa_2 a_0$ and the
dimensionless coupling constant $\lambda_-$ should be $\kappa_1/\kappa_2$.
Thus the dynamical length scale is $\xi\sim\exp-\kappa_2/\kappa_1$.
This conclusion was reached in Ref.~\onlinecite{White96} in the $({\rm
WZW}_{k=1})^2$ representation. We shall conclude from this that the fermions
have acquired a mass $m_i \sim v_i\xi^{-1}$. The first three fermions
($\psi_{1,2,3}$) have a common mass $m$, while $\psi_0$ has a slightly higher
mass $m_0$ (in absolute value), since $v_0>v_1=v_2=v_3$.
An additional velocity renormalization will take place during the RG flow,
but cannot be calculated by the above method.  In a diagrammatic technique, it
would show up at two loops, in a self-energy correction. The overall velocity
renormalization is important, since without it the long-distance theory would
have an $O(4)$ symmetry and all four fermions would have the same mass (up to
a sign).

In order to clarify the significance of these dynamically generated mass
scales, let us consider the following model:
\begin{equation}
\label{O(N)model}
{\cal L} = {1\over 2\pi}\sum_{i=1}^N v\,
(\psi_i\bar\partial\psi_i+\bar\psi_i\partial\bar\psi_i) +
\frac12\lambda\sum_{i\ne j} \psi_i\bar\psi_i \psi_j\bar\psi_j
\end{equation}
This model would be equivalent to
the Lagrangian (\ref{freeham},\ref{interac}) if all velocities were equal, if
$\lambda_+=0$ and if $N=4$, with $\psi_4=\psi_0$,
$\bar\psi_4=-\bar\psi_0$ (Kramers-Wannier duality). The model (\ref{O(N)model})
has $O(N)$ symmetry and a mass gap arises non-perturbatively in the spectrum if
$\lambda>0$. To see this in a mean-field approach, we assume that 
$\langle\psi_i\bar\psi_i\rangle=i\varepsilon\ne0$ (no sum over $i$) and
determine $\varepsilon$ self-consistently. Let us make the substitution
\begin{equation}
\psi_i\bar\psi_i \longrightarrow i\varepsilon + \psi_i\bar\psi_i
\end{equation}
in the Lagrangian, neglecting terms quartic in $\psi$, which
is equivalent to a large-$N$ approximation. We find the Lagrangian of $N$
massive fermions:
\begin{equation}
\label{O(4)hamil}
{\cal L} = \sum_{i=1}^N \bigg\{
{v\over 2\pi}\,(\psi_i\bar\partial\psi_i+\bar\psi_i\partial\bar\psi_i)
 + i\lambda(N-1)\varepsilon\psi_i\bar\psi_i\bigg\}
\end{equation}
where the mass is $m=2\pi\lambda(N-1)\varepsilon$. This mass may be
determined self-consistently, using the following expression for the Green's
function of real fermions:\cite{Itzykson}
\begin{mathletters}
\label{propag-psi}
\begin{eqnarray}\displaystyle
\langle \psi(0)\psi({\bf x})\rangle &=& \partial \int {d^2k\over\pi}\;
{e^{-i {\bf k}\cdot {\bf x}}\over {\bf k}^2+m^2}\\
\langle \bar\psi(0)\bar\psi({\bf x})\rangle &=& \bar\partial \int
{d^2k\over\pi}\; {e^{-i {\bf k}\cdot {\bf x}}\over {\bf k}^2+m^2}\\
\langle \psi(0)\bar\psi({\bf x})\rangle &=& -i\frac12 m \int {d^2k\over\pi}\;
{e^{-i {\bf k}\cdot {\bf x}}\over {\bf k}^2+m^2}
\end{eqnarray}
\end{mathletters}%
where ${\bf x}$ and ${\bf k}$ stand respectively for $(v\tau,x)$ and
$(i\omega/v,k)$. The mass $m$ is determined from the self-consistency condition
\begin{equation}
\label{selfcons}
{m\over 2\pi\lambda(N-1)} =
m\int{d^2k\over\pi}\; {e^{-i {\bf k}\cdot {\bf x}}\over {\bf k}^2+m^2}
\end{equation}
whose solution is, besides $m=0$,
\begin{equation}
m = \pm v\Lambda \exp-{1\over 2\pi\lambda(N-1)}
\end{equation}
where $\Lambda$ is a momentum cutoff. This solution exists only for positive
$\lambda$.

Returning to the Lagrangian (\ref{freeham},\ref{interac}) with
$\lambda_+$ renormalized to zero, all velocities equal and $\lambda=\lambda_-$,
this implies a mass gap $m\sim v\Lambda\exp(-1/6\pi \lambda_-)$, or 
$m\sim \kappa_2\exp(-\kappa_2/6\pi\kappa_1)$ if the characteristic velocity (of
order $\kappa_2$) is restored.
Since $(\psi_4,\bar\psi_4)=(\psi_0,-\bar\psi_0)$, the mass $m_0$ of the fourth
Ising model is equal to $-m$, if $v_0=v_i$. Since $v_0>v_i$, we conclude that
$-m_0 > m$.

A short remark about the sign of the mass: from the Ising model viewpoint, this
sign simply indicates on which side of the transition we stand: By convention,
$m>0$ in the disordered phase ($\langle\mu\rangle\ne0$) and $m<0$ in the
ordered phase ($\langle\sigma\rangle\ne0$). Of course, it is the absolute value
$|m|$ which occurs in the dispersion relation of the fermions.

The appearance of fermion mass terms breaks the diagonal ${\Bbb Z}_2$ symmetry
$(\psi_i,\bar\psi_i)\to(\psi_i,-\bar\psi_i)$ ($i=0-4$) of the full
Lagrangian (\ref{freeham},\ref{interac}). Thus, the ground state must be
doubly degenerate and the condensate $\langle\psi_i\bar\psi_i\rangle\ne0$
picks one of these ground states, the theory of massive fermions describing
excitations above that ground state only. This is consistent with the
Lieb-Schultz-Mattis theorem, which states that a half-integer spin chain with
local interactions and no explicit parity breaking has either no gap, or else
has degenerate ground states.

\section{Spin structure factor}
\label{correlS}
In a recent paper, Rao and Sen\cite{Rao96} have argued that dimerized spin
chains with second nearest-neighbor interactions admit possibly three different
phases (here we use the word `phase' to distinguish regions where the spin
structure factor $S(q)$ is not peaked at the same value of $q$). They name the
three phases as follows:\cite{noteC} a N\'eel phase ($S(q)$ is peaked at $\pi$),
a spiral phase ($S(q)$ is peaked at an intermediate momentum between $\pi$ and
$\pi/2$) and a collinear phase ($S(q)$ is peaked at $\pi/2$). In view of the
numerical results from Chitra and al.\cite{Chitra95}, the collinear phase
should not be stable for the spin-$\frac12$ chain. At first sight, there are
two paths that $S(q)$ may follow to go from the N\'eel phase to the collinear
phase. The first possibility is for the peak of $S(q)$ to move continuously
from $\pi$ to $\pi/2$, thus going through a spiral phase. The second
possibility is for the peak of $S(q)$ at $\pi$ to progressively decrease in
amplitude while a second peak at $\pi/2$ progressively appears; the system
might then go through a dimerized state. In view of this, the question of the
existence or not of a spiral phase for the frustrated spin-$\frac12$ chain
arises. To answer this question, we need to known how the spin structure factor
evolves as a function of the ratio
$\kappa_1/\kappa_2$. Unfortunately, in the present continuum approach we can
only calculate the spin-spin correlation function near $q=0$, $q=\pi$ and
$q=\pm\pi/2$. As seen below, this calculation is also interesting from the
point of view of symmetry and allows to relate the elementary spin
excitations to the fermions $\psi_i$.

The main conclusion of Sect.~\ref{NAbosS} is that the system (\ref{hamil1}) may
be described in the continuum limit by four noninteracting real fermions: three
with mass $m>0$ and velocity $v$, and one with mass $m_0<-m$ and velocity
$v_0>v$. The spin operator ${\bf S}_i$ is represented in terms of these
fermions through the relations (\ref{spinJJ}), (\ref{J:4fer})
and (\ref{g:4fer}). The $z$-component of the spin density has the following
expression near the wavevectors accessible to the continuum limit:
\begin{eqnarray}
\label{Szq}
S^z_{q\sim0} &\propto&~\psi_1\psi_2 +  \bar\psi_1\bar\psi_2\cr
S^z_{q\sim\pi} &\propto&~\psi_0\psi_3 +  \bar\psi_0\bar\psi_3\cr
S^z_{q\sim\pi/2} &\propto&~\mu_1\mu_2\sigma_3\sigma_0\cr
S^z_{q\sim-\pi/2} &\propto&~\sigma_1\sigma_2\mu_3\mu_0
\end{eqnarray}

Thus, the spin-spin correlation function near $q=0$
takes the form
\begin{equation}
\chi^{(0)} (x,\tau) \propto
\big\langle\left(\psi_1\psi_2+\bar\psi_1\bar\psi_2\right)(x,\tau)
\left(\psi_1\psi_2+\bar\psi_1\bar\psi_2\right)(0,0)\big\rangle
\end{equation}
while near $q=\pi$, it takes the following form:
\begin{equation}
\chi^{(\pi)}(x,\tau) \propto
\big\langle\left(\psi_3\psi_0+\bar\psi_3\bar\psi_0\right)(x,\tau)
\left(\psi_3\psi_0+\bar\psi_3\bar\psi_0\right)(0,0)\big\rangle
\end{equation}
In the first case ($q$ near 0), the two fermions have the same mass and
velocity, while in the second case ($q$ near $\pi$) they have different masses
and velocities. Consider the case of two fermions with different
masses ($m$ and $m'$) but identical velocities (for simplicity). The imaginary
part of the Fourier transform of the spin-spin correlation function -- i.e.,
the imaginary part of the dynamic susceptibility, or the spin structure factor
$S(q,\omega)$ -- may be calculated from the propagators (\ref{propag-psi}):
\begin{equation}
\label{chi1}
S(q,\omega)\propto{1\over u}\bigg[{q^2\over s^2}(m+m')^2+
{\omega^2\over s^2}(m-m')^2-{\omega^2+q^2\over s^4}(m+m')^2(m-m')^2\bigg]
\end{equation}
where $u$ and $s$ are defined by
\begin{eqnarray}
u^2&=&(s^2+m^2-m'^2)^2-4m^2s^2\\ s^2&=&\omega^2-v^2q^2
\end{eqnarray}
(in this expression we have returned to real frequencies).
In the case $m=m'$ this result becomes the spin structure factor near $q=0$:
\begin{equation}
S^{(0)}(q,\omega)\propto {m^2q^2\over s^3\sqrt{s^2-4m^2}}
\end{equation}
Neglecting velocity renormalization ($v_0=v_i$), the model is $O(4)$
symmetric at long distances and $m_0=-m$. Then the dynamical susceptibility
near $q=\pi$ would be given by the more general expression
(\ref{chi1}) with $m'=-m$. The expression of $S^{(\pi)}(q,\omega)$
appropriate for the more realistic case $v_0\ne v_i$ can be obtained in closed
form but is too cumbersome to display here.

The fact that $S^{(0)}(0,\omega)=0$ reflects the conservation of the total
magnetization. This is not the case for the total magnetization of each chain,
since $m\ne m'$, unless the two chains are decoupled ($m=m'=0$). Thus,
$S^{(\pi)}(0,\omega)\ne0$.

According to Eq.~(\ref{Szq}), the magnetic susceptibility near $q=\pm\pi/2$
is a product of four two-point functions of the Ising model, involving order
and disorder fields. For instance, near $q=\pi/2$,
\begin{eqnarray}
\label{Sqw:pi/2}
\chi^{(\pi/2)}(x,\tau) &\propto& \big\langle (\mu_1\mu_2\sigma_3\sigma_0)(0,0)
(\mu_1\mu_2\sigma_3\sigma_0)(x,\tau)\big\rangle\nonumber\\
&=& \tilde C^2(mr)C(mr)C(m_0r)
\end{eqnarray}
where $C(R)$ and $\tilde C(R)$ are respectively the two-point functions of the
order field and disorder field, as a function of the reduced distance
$R=\sqrt{x^2+v^2\tau^2}/\xi=mr$. These functions are known\cite{McCoy76} and
their leading asymptotic behavior is, in the disordered phase of the Ising
model,
\begin{equation}
C(R) = {A\over\pi} K_0(R) + O(e^{-3R})\qquad
\tilde C(R) = A\left\{ 1 + {1\over 8\pi R^2}e^{-2R} + O(e^{-4R})\right\}
\end{equation}
where $A$ is some constant and $K_{0,1}$ are the modified Bessel functions.
If the argument of $C$ is negative (i.e. for a negative mass), we perform
a Kramers-Wannier duality transformation and identify $C(-R)$ with 
$\tilde C(R)$. Thus, the leading asymptotic behavior of
the susceptibility near $q=\pi/2$ is
\begin{equation}
\chi^{(\pi/2)}(x,\tau) \propto \tilde C^2(mr)\tilde C(|m_0|r) C(mr)
\propto K_0(mr) + O(e^{-2mr})
\end{equation}
Notice that $K_0(R)$ is the real-space propagator of a free boson of
mass $m$. Thus, its Fourier transform is $\sim({\bf k}^2+m^2)^{-1}$.
Going back to real frequencies, the imaginary part of the susceptibility
has a pole at $\omega=\sqrt{(vk)^2+m^2}$, plus an incoherent part starting at
$\omega=2m$:
\begin{equation}
S^{(\pi/2)}(q,\omega) \propto {m\over|\omega|}\delta(\omega-\sqrt{(vk)^2+m^2})
+ \hbox{incoherent part}
\end{equation}
The magnetic susceptibility near $-\pi/2$ is obtained
by Kramers-Wannier duality:
\begin{equation}
\chi^{(-\pi/2)}(x,\tau) \propto \tilde C(mr) C^2(mr)C(|m_0|r)
\propto K_0(mr)^2 K_0(|m_0|r) + O(e^{-2mr})
\end{equation}
The associated spin structure factor has no single-particle peak, but instead
a continuum starting at $\omega=2m+|m_0|$. Thus, the single-particle magnetic
excitations live around $k=\pi/2$, whereas the excitations near $k=0,\pi$
have a two-particle behavior and those near $k=-\pi/2$ have a three-particle
behavior.

The above analysis assumed that the mass $m$ was positive, which amounts to
choosing one of the degenerate ground states, characterized by a short-range
order around $k=\pi/2$. If the other ground state were chosen, the mass $m$
would be negative and the above analysis could be repeated by interchanging the
roles of $k=\pi/2$ and $k=-\pi/2$. Thus, the spontaneous breakdown of parity is
reflected in the nonequivalence of $S(k,\omega)$ and $S(-k,\omega)$. However,
this would be unobservable in practice because of domain effects.

\section{String order parameter and ${\Bbb Z}_2\times {\Bbb Z}_2$
symmetry}
\label{stringS}
Kohmoto and Tasaki have shown\cite{Kohmoto92} that, for a spin-$\frac12$ chain
with dimerization, a string order parameter may be defined as den Nijs and
Rommelse have previously done for the spin-1 chain.\cite{Nijs89} A
nonlocal unitary transformation is introduced to show that the nonzero value of this
string order parameter is related to the breakdown of a ${\Bbb Z}_2\times {\Bbb Z}_2$ symmetry.
More recently, Shelton and al.\cite{Shelton96} showed that the string order
parameter:
\begin{equation}
O^z(n,m)=\exp\left\{ i\pi\sum_{j=n}^m\big(S_j^z+\tilde S_j^z\big)\right\}
\end{equation}
becomes, in the continuum limit,
\begin{equation}
\lim_{|x-y|\to\infty}\langle O^z(x,y)\rangle\sim
\langle\sigma_1\rangle^2\langle\sigma_2\rangle^2
-\langle\mu_1\rangle^2\langle\mu_2\rangle^2
\end{equation}
It was also argued in Ref.\onlinecite{Shelton96} that the nonzero value of this
order parameter is related to the breakdown of a ${\Bbb Z}_2\times {\Bbb Z}_2$
symmetry which, in the continuum limit, is given by the invariance under sign
inversion of both chiral components of each Majorana spinor: $\psi_a\rightarrow
-\psi_a$ and $\bar\psi_a\rightarrow -\bar\psi_a$ ($a$=1,2). This must be
accompanied by an inversion of both order and disorder fields:
$\sigma_a\to-\sigma_a$ and $\mu_a\to-\mu_a$.

Here, we expect a nonzero value of the string order parameter for two reasons.
First, since the $SU(2)$ symmetry cannot be spontaneously broken -- according
to the Mermin-Wagner theorem -- the mass of the first Ising model ($m_1$) must
be the same as that of the second Ising model ($m_2$), in order to keep the
symmetry under the exchange of the labels 1,2 and 3. It naturally implies that
they must have the same sign. So, if $\langle\sigma_1\rangle\neq 0$ ($m_1> 0$)
then $\langle\sigma_2\rangle\neq 0$ ($m_2> 0$). Similarly, if $\mu_1\neq 0$
($m_1< 0$) then $\langle\mu_2\rangle\neq 0$ ($m_2< 0$). Secondly, since a gap
open by the introduction of the interchain coupling, the masses $m_1$ and $m_2$
must be nonzero.

We can also reveal the presence of the ${\Bbb Z}_2\times {\Bbb Z}_2$ symmetry
without going to the continuum limit, i.e., directly from the Hamiltonian
(\ref{hamil1}). The unitary transformation $U$ introduced in
Ref.~\onlinecite{Kohmoto92} consists of many transformations applied in
succession. Explicitly, we have:
\begin{equation}
U=(D^\tau)^{-1}RDG
\end{equation}
where $G$ performs a rotation of $\pi$ about the $y$-axis on some
of the spins:
\begin{equation}
G=\bigotimes_{j=1}^{L/2}\exp\left[{i\pi\over2}(S_{4j-1}^y+S_{4j}^y)\right]
\end{equation}
$D$ is a duality transformation (see appendix A of reference\cite{Kohmoto92})
which introduces intersite spins. It is followed by a translation:
\begin{equation}
R:r\rightarrow{1\over 2}\left(r+\frac12\right)
\end{equation}
The spin on integer sites will be noted $\sigma$ and those on
the half-odd integer sites will be noted $\tau$. The final operation is to make an
inverse duality transformation for the $\tau$ spins. We refer the reader to the work
of Kohmoto and Tasaki\cite{Kohmoto92} for a full description of this unitary
transformation. If we apply this transformation to the Hamiltonian
(\ref{hamil1}), we find:
\begin{eqnarray}
UHU^{-1} =
\sum_{j=1}^N\bigg\{&\kappa_1&\big[\tau_j^x+\sigma_j^x-\sigma_j^x\tau_j^x-
\sigma_j^z\sigma_{j+1}^z-\tau_j^z\tau_{j+1}^z
-\sigma_j^z\tau_j^z\tau_{j+1}^z\sigma_{j+1}^z\big]\nonumber\\
+&\kappa_2&\big[\sigma_j^z\tau_j^x\sigma_{j+1}^z
+\sigma_j^y\tau_j^y\tau_{j+1}^z\sigma_{j+1}^z
+\tau_j^z\sigma_j^x\tau_{j+1}^z\nonumber\\ &&\qquad
+\sigma_j^z\tau_{j+1}^x\sigma_{j+1}^z
+i\sigma_j^z\tau_j^y\tau_{j+1}^y\sigma_{j+1}^y
+\tau_j^z\sigma_{j+1}^x\tau_{j+1}^z\big]\bigg\}
\end{eqnarray}

The $\sigma$'s and $\tau$'s are sets of Pauli matrices. The new Hamiltonian
$\tilde H=UHU^{-1}$ is clearly invariant under a rotation of $\pi$ about the
$x$ axis applied to the $\sigma$-spins alone or the $\tau$-spins alone. An
four-fold degeneracy of the ground state of $H$ in the thermodynamic limit does
not follow from this broken symmetry since this is not a local
symmetry.\cite{Kennedy92}

\section{Discussion}
\label{discussionS}

The crucial difference between the Hamiltonian (\ref{hamil1}) and that of the
more familiar spin ladder is the occurrence, in the latter, of an
interaction term of the form
\begin{equation}
\label{ladder1}
{\cal L}_{\rm ladder} ={\eta\over2\pi} {\rm Tr}(g\tau^a){\rm Tr}(g'\tau^a)
\end{equation}
($\eta$ is some constant proportional to the interchain coupling).
Since the matrix field $g$ has conformal dimensions $(\frac14,\frac14)$, the
above perturbation has scaling dimension $1$: it is relevant. If the
representation (\ref{g:4fer}) and the OPE's (\ref{OPE:Ising}a-d) are used to
express this interaction in terms of fermions, one finds
\begin{equation}
\label{ladder2}
{\cal L}_{\rm ladder} =
i{\eta\over2\pi}
(\psi_1\bar\psi_1+\psi_2\bar\psi_2+\psi_3\bar\psi_3-3\psi_0\bar\psi_0)
\end{equation}
This coincides with the conclusions of Ref.~\onlinecite{Shelton96}, obtained
by Abelian bosonization. The mass terms now appear explicitly, with a triplet
of mass $\eta$ and a singlet of mass $-3\eta$. The interchain coupling
explicitly breaks the invariance under parity that is spontaneously broken in
the `zig-zag' case. If the two rungs of the zig-zag had different couplings
($\kappa_1$ and $\kappa_1'$), an interaction like (\ref{ladder1}) would be
generated and the gap would have a linear dependence on the interchain
coupling. Of course, the marginal interaction (\ref{interac}) is always present
and provides an additional renormalization of the masses. As
$\kappa'_1\to\kappa_1$, the dependence of the gap on the interchain coupling
should become more and more exponential because of this renormalization.

We were concerned in this work with the regime $\kappa_1\ll\kappa_2$ and the
conclusions are nominally valid only in this regime, although we expect them
to be qualitatively correct even for $\kappa_1\sim\kappa_2$. However, in the
opposite regime ($\kappa_1\gg\kappa_2$) the system should be treated as a
single chain and we should perturb around a single WZW$_{k=1}$ model. This is
explained in Ref.~\onlinecite{Affleck89}. The conclusion is that the
perturbation is marginally irrelevant if the ratio $\kappa_2/\kappa_1$ is
smaller than some critical value, and leads to an
exponential gap above that critical value. In that regime the ground state is
spontaneously dimerized (spontaneous breaking of parity). This conclusion is
also valid in the regime $\kappa_1\ll\kappa_2$. Indeed, the order parameter for
dimerization
\begin{equation}
d = \langle {\bf S}_{2i}\cdot {\bf S}_{2i-1}-{\bf S}_{2i}\cdot {\bf S}_{2i+1}
\rangle
\end{equation}
coincides, in the continuum limit, with the ladder perturbation
(\ref{ladder1}), up to terms that have a vanishing expectation value.
Translated in terms of the bare interaction couplings $\lambda_\pm$ of
Eq.~(\ref{interac2}) and of the masses $m,m_0$, the spontaneous dimerization
becomes
\begin{equation}
d \propto m_0\left[ m(\lambda_- - \lambda_+) + m_0(\lambda_- +
\lambda_+)\right]
\end{equation}
This is generically nonzero.

So far we have supposed that $\kappa_1$ is positive, corresponding to an
antiferromagnetic interchain coupling. The ferromagnetic case may be treated
just as well. In that case, both interaction constants
$\lambda_\pm$ of Eq.~(\ref{interac2}) are negative and thus renormalize to
zero: the model is equivalent to a theory of four free Majorana fermions,
with different velocities. From Eq.~(\ref{vrenorm2}) with negative $\lambda_1$,
we expect the velocity $v_0$ of $\psi_0$ to be smaller than the velocity $v$ of
$\psi_{1,2,3}$. Thus, we conjecture that the ferromagnetic model is critical,
albeit with two sectors having different velocities: a triplet sector
equivalent to the WZW$_{k=2}$ theory and a singlet sector with a smaller
velocity. This is not the same as saying that the two chains
{\it decouple} at long distances, since in that case the structure of
excitations would be different. This conjecture might be tested by exact
diagonalizations on finite systems and some information on the velocity
renormalization might be extracted this way.

\acknowledgements 
Discussions with P.~Mathieu and P.~Di~Francesco are gratefully acknowledged.
This work is supported by NSERC (Canada) and by F.C.A.R. (le
Fonds pour la Formation de Chercheurs et l'Aide \`a la Recherche du
Gouvernement du Qu\'ebec).

\appendix
\section{WZW models}
A systematic review of WZW models cannot take place in a regular paper.
Here we simply recall basic concepts and a few definitions, in order to
fix the notation and the normalization used in this work. We follow in this
respect Ref.~\onlinecite{CFT96}.

Wess-Zumino-Witten (WZW) models are defined in terms of a matrix-valued
field $g$ belonging to a unitary representation of $SU(2)$ (more generally,
of a Lie group 
${\frak g}$) with the following action:\cite{Witten84,Knizhnik84}
\begin{equation}
\label{WZWaction}
S = {k\over16\pi}\int d^2x~ {\rm Tr}'(\partial^\mu g^{-1}\partial_\mu g)
- {ik\over 24\pi}\int_B d^3y~\varepsilon^{\mu\nu\rho}
{\rm Tr}'(g^{-1}\partial_\mu g g^{-1}\partial_\nu g g^{-1}\partial_\rho g)
\end{equation}
where the trace ${\rm Tr}'$ is proportional to the usual trace operation:
\begin{equation}
{\rm Tr}' = {1\over x_s} {\rm Tr} \qquad x_s = \frac13 s(s+1)(2s+1)
\end{equation}
($s$ is the spin of the representation). $k$ is a positive integer called
the {\it level} of the WZW model.
The first term of (\ref{WZWaction}) is the usual nonlinear sigma model. The
second term is topological and is integrated on a three-dimensional
manifold $B$ of which two-dimensional space-time is the boundary. Its value
is independent of the precise form of $B$ (modulo $2\pi$), provided $k$ is an
integer. 

The fundamental property of the WZW model -- enforced by the
relative normalization of the two terms of the action (\ref{WZWaction}) -- 
is its full conformal symmetry. For this reason, it is best
described in the language of conformal field theory, with holomorphic
(or left) and antiholomorphic (or right) coordinates
\begin{eqnarray}
z &= -i(x-vt) &= v\tau-ix \cr \bar z &= i(x+vt) &= v\tau+ix
\end{eqnarray}
where $\tau=it$ is the Euclidian time and $v$ is the characteristic velocity
of the model, implicit in the covariant notation of Eq. (\ref{WZWaction}).
The left and right derivatives are commonly used:
\begin{equation}
\label{derivs}
\partial \equiv \partial_z = \frac12\left({\partial\over\partial x} 
+ i{1\over v}{\partial\over\partial \tau}\right)
\qquad
\bar\partial \equiv \partial_{\bar z} = \frac12\left({\partial\over\partial x} 
- i{1\over v}{\partial\over\partial \tau}\right)
\end{equation}
The WZW model has $SU(2)$ symmetry and this entails the existence of a
conserved current $J_\mu$, expressed here in its left ($z$) and
right ($\bar z$) components:
\begin{equation}
J \equiv J_z = \partial g g^{-1} \qquad
\bar J \equiv J_{\bar z} = g^{-1}\bar\partial g
\end{equation}
Closely related to its conformal symmetry is the separate conservation
of the left and right currents: $\partial\bar J=0$ and
$\bar\partial J=0$ (the $SU(2)$ symmetry is enlarged to a chiral
symmetry $SU(2)_L\otimes SU(2)_R$). Hence $J(z)$ depends only on $z$ and
$\bar J(\bar z)$ on the $\bar z$. These matrix currents may be decomposed
along a basis of spin-$s$ generators. For spin-$\frac12$, we choose
\begin{equation}
J(z) = J^a(z)\tau^a \qquad \bar J(\bar z) = \bar J^a(\bar z)\tau^a
\end{equation}
where the $\tau^a$ are the usual Pauli matrices.

In practice, the action (\ref{WZWaction}) is not useful for practical
calculations. The traceless, symmetric energy-momentum tensor, which generates
local conformal transformations (in particular space-time translations) is
more useful. Its two nonzero components are given by the so-called Sugawara
form:
\begin{equation}
T(z) = {1\over (k+2)}(J^aJ^a)\qquad
\bar T(\bar z) = {1\over (k+2)}(\bar J^a \bar J^a)
\end{equation}
The notation $\big(\ldots\big)$ above stands for a normal ordering
(regularized product). The dynamics of the theory is determined by the
short-distance product (operator-product expansion, or OPE) of the various
fields. The OPE of $T(z)$ with a local scaling (or {\it primary}) field
$\phi(w,\bar w)$ reflects the conformal (or scaling) properties of that
field:
\begin{equation}
\label{OPE:T-phi}
T(z)\phi(w) \sim {h\phi(w,\bar w)\over(z-w)^2} + 
{\partial_w\phi(w,\bar w)\over z-w} 
\end{equation}
where $h$ is the conformal dimension of the field $\phi$ and the symbol
$\sim$ means an equality modulo terms which are regular as $z\to w$. A similar
expression holds for $\bar T$ and the sum $h+\bar h$ is the usual scaling
dimension. The OPE of $T$ with itself is slightly different:
\begin{equation}
\label{OPE:TT}
T(z)T(w) \sim {c/2\over(z-w)^4} + {2T(w)\over(z-w)^2} + 
{\partial_wT(w)\over z-w}
\end{equation}
The constant $c$ in the most singular term is the {\it central charge} of the
conformal theory and measures the number of degrees of freedom of the
theory; its value in the $SU(2)$ WZW model is
\begin{equation}
c = {3k\over k+2}
\end{equation}

The OPE of the currents $J$ and $\bar J$ with a local matrix field $\phi$
fields reflects its transformation properties under the action of $SU(2)$:
\begin{equation}
\label{OPE:J-g}
J^a(z)g(w,\bar w) \sim -\frac12{\tau^a g(w,\bar w)\over z-w}  \qquad
\bar J^a(\bar z)g(w,\bar w) \sim \frac12{g(w,\bar w)\tau^a\over z-w} 
\end{equation}
The OPE of the current with itself constitutes the so-called {\it current
algebra}:
\begin{eqnarray}
\label{OPE:JJ}
J^a(z)J^b(w) &\sim& {(k/2)\delta_{ab}\over(z-w)^2} + 
i \varepsilon_{abc}{J^c(w)\over z-w} \nonumber\\
\bar J^a(\bar z)\bar J^b(\bar w) &\sim& {(k/2)\delta_{ab}\over(\bar z-\bar
w)^2} +  i \varepsilon_{abc}{\bar J^c(\bar w)\over \bar z-\bar w} \nonumber\\
J^a(z)\bar J^b(\bar w) &\sim& 0
\end{eqnarray}

The WZW at level $k$ contains several matrix-valued scaling fields,
one for each value of the spin $s$ up to (and including) $s=k/2$. The
conformal dimensions $h$ and $\bar h$ of the spin-$s$ field are
\begin{equation}
h_s = \bar h_s = {s(s+1)\over k+2}
\end{equation}
The OPE of the various matrix fields is governed by the rule of addition of
angular momenta and by the constraint that no field of spin $s>k/2$ occurs
in the operator algebra. These OPE's were calculated in
Ref.~\onlinecite{Fateev86}. We shall only be concerned with the simplest
case ($k=1$).

The level-1 WZW model has central charge $c=1$ and contains a single matrix
field $g_{n\bar n}$ ($n,\bar n=\pm\frac12$) of conformal dimensions
$(\frac14,\frac14)$. With the proper normalization, its OPE is
\begin{equation}
\label{OPE:k=1}
g_{n\bar n}(z,\bar z)g_{m\bar m}(w,\bar w) \sim 
{1\over|z-w|}\varepsilon_{nm}\varepsilon_{\bar n\bar m}
\end{equation}
where $\varepsilon_{nm}$ is the antisymmetric symbol.
We may use the following decomposition:
\begin{equation}
\label{decomp}
g(z,\bar z) = \frac12\sum_{i=0}^3 g_a(z,\bar z)\tau^a
\qquad g_a = {\rm Tr}(\tau^a g)
\end{equation}
The OPE of $g$ with itself is then
\begin{equation}
\label{OPE:gg}
g_a(z,\bar z)g_b(w,\bar w)\sim {2\delta_{ab}\over |z-w|}(-1)^{\delta_{a0}+1}
\end{equation}
%

\section{The Ising model}
It is well known that the two-dimensional Ising model is equivalent to a real
(Majorana) fermion in 1+1 dimension. The critical point of the Ising model
corresponds to the massless point of the fermion theory and constitutes one
of the simplest conformal field theories, of central charge $c=\frac12$. This
theory contains a two-component fermion $(\psi(z),\bar\psi(\bar z))$.
The holomorphic field $\psi$ has conformal dimensions $(\frac12,0)$ while
its antiholomorphic counterpart $\bar\psi$ has conformal dimensions
$(0,\frac12)$. Their OPE is
\begin{eqnarray}
\label{OPE:fermion}
\psi(z)\psi(w)&\sim&{1\over z-w}\cr
\bar\psi(\bar z)\bar\psi(\bar w)&\sim&{1\over \bar z-\bar w}\cr
\psi(z)\bar\psi(\bar w)&\sim& 0
\end{eqnarray}
The product $\varepsilon=i\bar\psi\psi$ has conformal dimensions 
$(\frac12,\frac12)$ and is called the {\it energy} field; it is the mass
term that takes the model away from its critical point.
The energy-momentum tensor of the fermion theory is
\begin{equation}
\label{EMtensor}
T(z) = -\frac12\psi\partial\psi \qquad
\bar T(\bar z) = -\frac12\bar\psi\bar\partial\bar\psi
\end{equation}

The critical Ising model also contains an {\it order} field 
$\sigma(z,\bar z)$ which is the continuum limit of the Ising spin. This
field has conformal dimensions $(\frac1{16},\frac1{16})$ and is not locally
related to the fermion field. Indeed, in the transfer-matrix description of
the 2D Ising model, the fermion field is introduced by a (nonlocal)
Wigner-Jordan transformation. The Kramers-Wannier duality transformation
of the Ising model maps the order field $\sigma$ into a disorder field
$\mu$ which has essentially the same properties, except that
$\langle\sigma\rangle\ne0$,
$\langle\mu\rangle=0$ in the ordered phase and $\langle\sigma\rangle=0$,
$\langle\mu\rangle\ne0$ in the disordered phase. At the (massless) critical
point, both fields have a vanishing expectation value. All three fields
($\psi,\sigma,\mu$) are mutually nonlocal, which is reflected in their OPE by
the existence of branch cuts. These OPE's are given below:
\begin{mathletters}
\label{OPE:Ising}
\begin{eqnarray}
&&\sigma(z,\bar z)\sigma(w,\bar w) \sim 
{1\over|z-w|^{1/4}} + \frac12|z-w|^{3/4}\varepsilon(w,\bar w) \\
&&\mu(z,\bar z)\mu(w,\bar w) \sim 
{1\over|z-w|^{1/4}} - \frac12|z-w|^{3/4}\varepsilon(w,\bar w) \\
&&\sigma(z,\bar z)\mu(w,\bar w) \sim 
{\gamma(z-w)^{1/2}\psi(w)+\beta\gamma^*(\bar z-\bar w)^{1/2}\bar\psi(\bar w)
\over\sqrt2 |z-w|^{1/4}}\\
&&\mu(z,\bar z)\sigma(w,\bar w) \sim
{\gamma^*(z-w)^{1/2}\psi(w)+\beta\gamma(\bar z-\bar w)^{1/2}\bar\psi(\bar w)
\over\sqrt2 |z-w|^{1/4}}\\
&&\psi(z)\sigma(w,\bar w) \sim 
{\gamma\mu(w,\bar w)\over\sqrt2 (z-w)^{1/2}}\\
&&\psi(z)\mu(w,\bar w) \sim 
{\gamma^*\sigma(w,\bar w)\over\sqrt2 (z-w)^{1/2}}\\
&&\bar\psi(\bar z)\sigma(w,\bar w) \sim 
{\beta\gamma^*\mu(w,\bar w)\over\sqrt2 (\bar z-\bar w)^{1/2}}\\
&&\bar\psi(\bar z)\mu(w,\bar w) \sim 
{\beta\gamma\sigma(w,\bar w)\over\sqrt2 (\bar z-\bar w)^{1/2}}
\end{eqnarray}
\end{mathletters}%
where some arbitrariness remains in the constants $\beta$ and $\gamma$
because of the nonlocal character of these OPE's: $\beta=\pm1$ and
$\gamma=\pm e^{\pm i\pi/4}$. In this work we choose $\gamma=e^{i\pi/4}$
and $\beta=1$.

Since the regularized product of $\sigma$ with $\mu$ is a fermion, these
operators must carry some anticommuting character. We shall assume that
the disorder operator $\mu$ anticommutes with all fermion fields and other
disorder operators, but not with itself nor with the order fields. This is
a matter of convention ($\sigma$ could have been chosen instead).

%

\begin{figure}[tpb]
\vglue 0.4cm\epsfxsize 8cm\centerline{\epsfbox{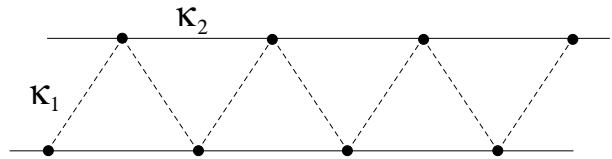}}\vglue 0.4cm
\caption{
The `zig-zag' chain, with interchain coupling $\kappa_1$ and intrachain
coupling $\kappa_2$, also equivalent to a single chain with NN coupling
$\kappa_1$ and NNN coupling $\kappa_2$.}
\label{fig1}
\end{figure}


\begin{references}
%
\bibitem[*]{noteA}
Unless said otherwise, we will mean by `WZW model' the $SU(2)$ WZW
model, at a specified level $k$.

\bibitem[\dagger]{noteB}
Note that the expression (\ref{spinJJ}) cannot be substituted into
(\ref{Heisenberg}) to find the WZW model Hamiltonian! This incorrect procedure
yields the wrong sign for the marginal perturbation (\ref{marginal1}). 
Eq.~(\ref{spinJJ}) should be used only to evaluate correlation functions or
express perturbations added to the half-filled Hubbard model.

\bibitem[\ddagger]{noteC}
In this section, the system will be  regarded as just one chain with NNN
interactions and not as two chains with a zig-zag interaction.

\bibitem{Haldane83}
F. D. M. Haldane, \pl{93A}, 464 (1983); \prl{50}, 1153 (1983).

\bibitem{Affleck86}
I. Affleck, Nucl. Phys. B{\bf 265}, 409 (1986). 

\bibitem{Affleck89} I. Affleck. in {\it Les Houches, session XLIX, 1988,
Champs, Cordes et Ph\'enom\`enes Critiques} (Elsevier, New York, 1989).

\bibitem{Witten84}
E. Witten, Comm. Math. Phys. {\bf 92}, 455 (1984).

\bibitem{Matsuda95}
M. Matsuda and K. Katsumata, J. Mag. Mag. Mat. {\bf 140-145}, 1671 (1995).

\bibitem{White96}
S. R. White and I. Affleck, preprint (cond-mat 9602126).

\bibitem{Nijs89}
M. den Nijs and K. Rommelse, Phys. Rev. B~{\bf 40}, 4709 (1989).

\bibitem{Totsuka95}
K. Totsuka and M. Suzuki, J. Phys. Condens. Matter {\bf 7}, 6079 (1995).

\bibitem{Fateev86}
A. B. Zamolodchikov and V. A. Fateev, Sov. J. Nucl. Phys. {\bf 43}, 657 (1986).

\bibitem{Shelton96}
D. G. Shelton, A. A. Nersesyan and A. M. Tsvelik, Phys. Rev. B~{\bf 53}, 8521
(1996).

\bibitem{Polyakov72}
A. M. Polyakov, Zh.ETF {\bf 63}, 24 (1972).

\bibitem{Itzykson}
C.~Itzykson and J.-M.~Drouffe, {\it Statistical Field Theory}, Cambridge
University Press, 1989; {\it Th\'eorie Statistique des Champs},
Inter\'Editions/\'Editions du C.N.R.S., 1989.

\bibitem{Rao96}
S. Rao and D. Sen, preprint (cond-mat 9604044).

\bibitem{Chitra95}
R. Chitra, S. Pati, H. R. Krishnamurthy, D. Sen, S. Ramasesha,
Phys. Rev.~B~{\bf 52}, 6581 (1995).

\bibitem{McCoy76}
T.T. Wu, B. McCoy, C.A.Tracy and E. Barouch, Phys. Rev. B~{\bf 13}, 316 (1976).
See also Ref.~\onlinecite{Itzykson} above.

\bibitem{Kohmoto92}
M. Kohmoto and H. Tasaki, Phys. Rev. B~{\bf 46}, 3486 (1992).

\bibitem{Kennedy92}
T. Kennedy and H. Tasaki, Phys. Rev. B~{\bf 45}, 304 (1992).

\bibitem{CFT96}
P. Di~Francesco, P. Mathieu and D. S\'en\'echal,
{\it Conformal Field Theory}, Springer-Verlag (in press).

\bibitem{Knizhnik84}
V. G. Knizhnik and A. B. Zamolodchikov, Nucl. Phys. B{\bf 247}, 83 (1984).

\end{references}
\end{document}